\documentstyle[11pt,aussois2003,epsf,twoside]{article}
\def\edcomment#1{\iffalse\marginpar{\raggedright\sl#1\/}\else\relax\fi}
\reversemarginpar

\begin{document}

\title{Recent Microlensing Results from the MACHO Project}

\author{P.~Popowski$^1$, C.A.~Nelson, D.P.~Bennett, A.J.~Drake,
T.~Vandehei, K.~Griest, K.H.~Cook, C.~Alcock, R.A.~Allsman,
D.R.~Alves, T.S.~Axelrod, A.C.~Becker, K.C.~Freeman, M.~Geha, 
M.J.~Lehner, S.L.~Marshall, D.~Minniti, B.A.~Peterson, 
P.J.~Quinn, C.W.~Stubbs, W.~Sutherland,
D.~Welch (The~MACHO~Collaboration)}
\affil{$^1$ Max-Planck-Institut f\"{u}r Astrophysik,
Karl-Schwarzschild-Str. 1, 85741 Garching, Germany; e-mail: popowski@mpa-garching.mpg.de}

\label{page:first}

\begin{abstract}
We describe a few recent microlensing results from the MACHO
Collaboration. 
The aim of the MACHO Project was the identification and quantitative
description of dark and luminous matter in the Milky Way using
microlensing toward the Magellanic Clouds (LMC and SMC) and the
Galactic bulge. 
We start with a discussion of the Hubble Space
Telescope follow-up
observations of the microlensing events toward the LMC detected in the 
first 5 years of the experiment. Using
color-magnitude diagrams we attempt to distinguish between two possible
locations of the microlensing sources toward the LMC: 1) in the LMC or
2) behind the LMC. 
We conclude that unless the extinction is
extremely patchy, it is very unlikely that most of the
LMC microlensing events have source stars behind the LMC.
During examination of the HST images of the 13 LMC events we found a very red
object next to
the source star of event LMC-5. Based on astrometry, microlensing parallax fit
and a spectrum, we argue that in this case
we directly image the lens -- a low-mass disk star. 

Then we focus on the majority of the events observed by the MACHO
Project, which are detected toward
the Galactic bulge. 
We determine the microlensing optical depth, which describes
the amount of matter between us and the Galactic center.
We argue that the microlensing optical depth toward the bulge
is best measured using only a subclass of the events, namely the ones 
that have clump giant sources. 
They are numerous and belong to the 
brightest stars
in the bulge, which makes them insensitive to blending bias.
Our analysis of those events suggests that the optical depth toward
the Galactic bulge is 
$\tau_{\rm bulge} = (1.4 \pm 0.3) \times 10^{-6}$, in good
agreement with other observational constraints and with theoretical
models.
There are many
long-duration events among the bulge candidates. We take advantage
of this situation investigating the microlensing parallax effect.
We show that the events with the strongest parallax signal are probably
due to massive remnants. Events MACHO-96-BLG-5
and MACHO-98-BLG-6 might have been caused
by the black holes with masses of order of $6$ solar masses.

\end{abstract}

\section{The MACHO Microlensing Experiment}
 
The Milky Way is one of the two most massive galaxies in the Local Group
and so is likely to represent a typical product of cosmological evolution
of matter trapped in a relatively deep potential well.
The detailed formation processes of our Galaxy put constraints on
cosmology, star formation history, environmental influence on galactic
evolution etc.
The key to understanding the formation process of the Galaxy is to understand
its structure and stellar populations.
There are still many questions related to the structure of the Galaxy
that have not been satisfactory answered.
``What is the Galactic dark halo made off?'' is probably the most
momentous question but the others related to the inner structure of our
Galaxy or the origin of different components also have profound
consequences for our understanding of the galaxy formation.

Microlensing is one of the preferred techniques to study the Milky Way
because of its sensitivity to both luminous and dark matter.
Microlensing results from the bending of light by massive objects.
The signal depends sensitively on the alignment between the observer,
the source of light (at a distance $D_s$), and a lens located between them
(at a distance $D_l$). Strong signal is observed if at some time
the projected distance between the source and the lens is of order
of the Einstein radius, $r_E$, defined as:
\begin{equation}
r_E = \sqrt{\frac{4G m_{\rm lens}}{c^2} \frac{(D_s-D_l)D_l}{D_s}}, \label{re}
\end{equation}
and the event produces significant magnification on a time scale
given by
\begin{equation}
t_E = \frac{r_E}{v_{\perp}}, \label{te}
\end{equation}
where $m_{\rm lens}$ is the mass of the lens, and $v_{\perp}$ a relative
transverse velocity.
The observable quantity, Einstein radius crossing time, $t_E$,
is a degenerate combination of $m_{\rm lens}$, $v_{\perp}$, $D_{l}$
and $D_s$. Fortunately, 
microlensing phenomenology goes much beyond the simple point
lens approximation. 
The effects of 'parallax' 
(Gould 1992; Alcock et al.\ 1995), 
binary caustic crossing (Mao \& Paczy\'{n}ski 1991; Afonso et al. 2000), 
or finite source (Gould 1994; Alcock et al.\ 1997b) 
allow one to break degeneracies 
present in the simplest cases and provide useful constraints on stellar 
physics.
Below we will use the effect of parallax, from which one may infer
the projected transverse velocity:
\begin{equation}
\tilde{v} = v_{\perp} \frac{D_s}{D_s-D_l}. \label{tildev}
\end{equation}
Substitution of equations (2) %(\ref{tildev}) 
and (3) %(\ref{te}) 
into (1) %(\ref{re}) 
results in:
\begin{equation}
M = \frac{c^2}{4G} t_E^2 {\tilde{v}}^2 \frac{D_s-D_l}{D_s D_l} \equiv 
\frac{c^2}{4G} t_E^2 \tilde{v} \mu_{\rm rel}, \label{mass}
\end{equation}
where in the second equity we simply recognized the fact that
the relative proper motion of the lens and the source was equal to
$\mu_{\rm rel}= v_{\perp}/D_l$.
When we have $\tilde{v}$ from the parallax fit to the
microlensing light curve, then the only uncertainty in the mass
comes from the lack of knowledge of the distance to the lens
(distance to the source is oftentimes approximately known). With the 
additional measurement
of the relative proper motion, one solves for the mass with
no ambiguity.

The amount of mass between the source and observer and its
distribution along the line of sight is described by
the optical depth, which is the probability that microlensing
happens on a given star (typically of order $10^{-7}-10^{-6}$). 
The actual count of events (microlensing
rate) is proportional to the optical depth and to the number 
of sources. Therefore, microlensing surveys target concentrations
of stars to maximize the number of detected events.
Microlensing observations toward the LMC and SMC test the existence and 
composition of the lensing population in the Galactic dark halo.
The microlensing optical depth toward
the bulge is a function of the matter content of the inner Galaxy
and is sensitive to the characteristics of the Galactic model.

\begin{figure}
\plotfiddle{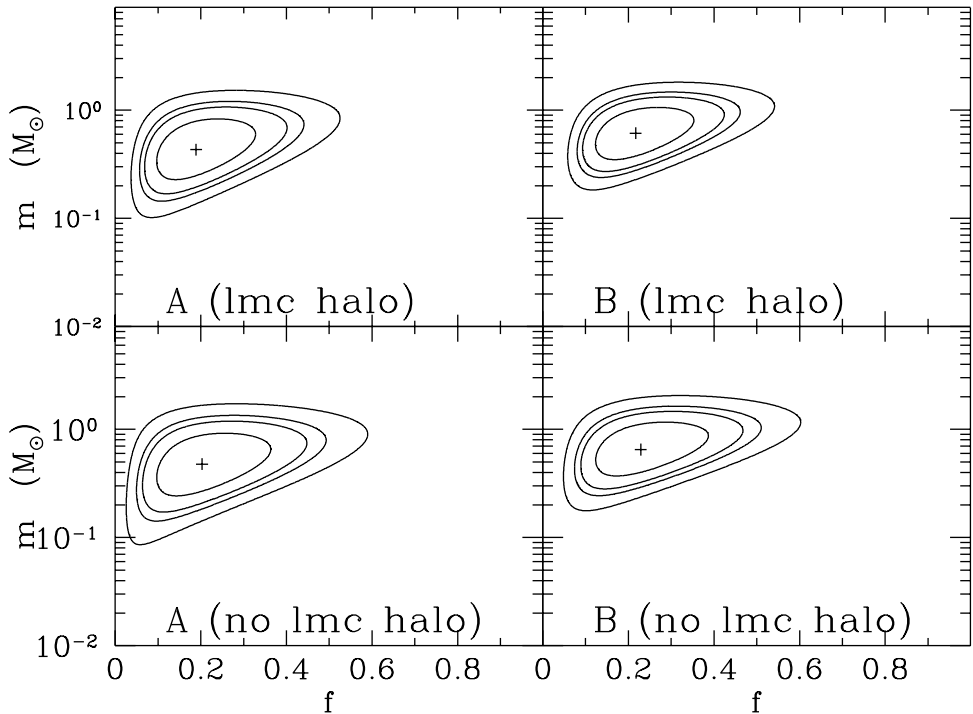}{4.5in}{-90}{50.0}{50.0}{-200}{300}
\caption{Alcock et al.\ (2000b) LMC results for the model of standard
spherical halo. The plots show the most likely mass of a typical
lens $m$ versus
the fraction $f$ of the dark halo in MACHOs. The four cases correspond 
to different cuts and assumptions about the LMC structure.}
\end{figure}

Less than two decades ago,
microlensing was still at a stage of a purely theoretical concept
(Paczy\'{n}ski 1986, 1991; Griest 1991). It became
such a powerful astrophysical tool thanks to the experiments like the MACHO
Survey, which had a major impact on the development and
application of the microlensing technique to Galactic studies.
The MACHO experiment collected images of the Galactic bulge and
Magellanic Clouds from 1992 through 1999. All observations were taken with
the 1.3~m Great Melbourne Telescope with a dual-color wide-field camera.
The MACHO camera consisted of two sets of four 2k x 2k CCDs that
collected blue ($B_M$) and red ($R_M$) images simultaneously using
a dichroic beam splitter. A single observed field covered an area of
43' by 43'. We observed 82 fields in the LMC, 21 in the SMC and 
94 in the Galactic bulge region, and collected about 100000 images. 
The fields with high priority were
observed on most photometric nights in the season. 
The LMC observations were carried out all year long, whereas the
bulge season lasted from March till October.
The resulting light curves often
contain several hundred points with individual fields sampled on
average every few days.
Details of the MACHO telescope system are given by Hart et al.\ (1996)
and of the camera system by Stubbs et al.\ (1993) and Marshall et al.\ (1994). 
Details of the MACHO imaging, data reduction and photometric calibration 
are described by Alcock et al.\ (1999).

\section{Where are the source stars of the LMC events?}

We (Alcock et al.\ 2000b) used two sets of microlensing candidates:
conservative set A of 13 events and inclusive set B of 17 events
to argue that Massive Compact Halo Objects (MACHOs) contribute of
order of 20\% to the mass of the dark halo of the Milky Way.
Using maximum likelihood technique, we showed that the most likely mass
of a typical lens is $\sim 0.5 M_{\sun}$.
Upper limits on the halo contribution of MACHOs consistent with our
results were also obtained by the
EROS collaboration (Lasserre et
al.\ 2000) based on 3 events in the LMC or 4 events in
both Magellanic Clouds.
Neither of the groups had enough statistical power to
reveal the nature or the location of the lenses along the line of
sight. Figure 1 summarizes the MACHO results for a standard model
of the Galactic dark halo.

\begin{figure}
\plotone{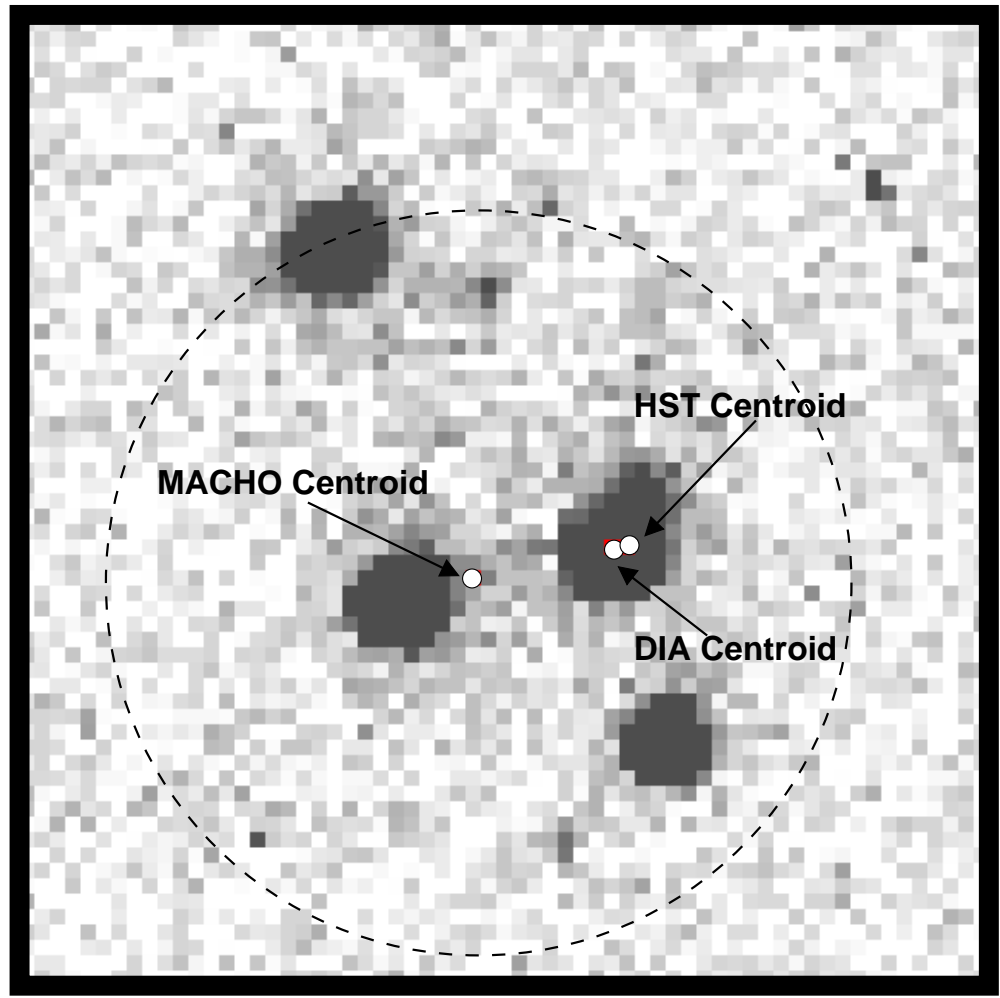}
\caption{Illustration of the identification of a microlensing source
star with Difference Image Analysis. The circle traced with a dotted
line is a typical MACHO seeing disk. The centroid of MACHO light,
the centroid of the event source star as inferred from DIA analysis,
and the HST centroid of the star closest to this source are marked.}
\end{figure}

In all cases when we have some additional constraints on the location
of the sources, they seem to belong to the known stellar systems
(Afonso et al.\ 2000; Alcock et al.\ 2001a; Alcock et al.\ 2001c), 
which makes some to believe that all of
the events must be of this nature. This probability argument is 
however faulty to some extent. Most effects that give the
indication of the lens position occur preferentially when the
lens is close to either observer (parallax) or the source (xallarap).
Therefore, they provide little information about the location
or population membership of the remaining events.

Being unable to conclusively decide on the location of the lenses
from available microlensing information, we will now attempt to obtain
some indirect bounds. The idea we want to entertain here is the
following: find the location of the sources first and then infer
the location of the lenses.
A useful demonstration of how this works would be to consider two
situations with somewhat oversimplified conclusions:
1) sources are in the LMC $\rightarrow$ lenses are in the Milky Way,
2) sources are behind the LMC
$\rightarrow$ lenses are in the LMC.
The standard case 2) was considered by Zhao, Graff, \& Guhathakurta (2000), who
placed the source population $\sim 7$ {\rm kpc} behind the main body of the LMC
and assumed uniform LMC extinction of $E(B-V)=0.13$.
We will not analyze all the possible locations of the
sources and the implications for the lens population here, 
but we note that this issue is discussed in detail by
Alcock et al.\ (2001b) and Nelson et al.\ (2003).

The sky appearance at the MACHO observing site at Mount Stromlo is
very similar to the one portrayed
in ``Starry Night'', the oil canvas painted by Vincent van Gogh
in Saint--R\'{e}my in 1889\footnote{Color reproduction available at
{\tt http://www.vangoghgallery.com/painting/p\_0612.htm}.}.
With median seeing above 2 arc-seconds, the individual stars in the
LMC are typically not resolved. The objects observed are blends of
several stars of different brightness (Alcock et al.\ 2001b,d).
Therefore, we use Hubble Space Telescope (HST) images to construct 
true color-magnitude diagrams (CMDs) and Difference
Image Analysis (DIA) to establish magnitudes and colors of our source stars.
The HST and MACHO frames are put on the same astrometric system,
and then the centroid of the source star is recovered based on the
motion of the centroid of light during the event. 
The HST star closest to the recovered centroid is assumed to be the
source star.
This procedure is illustrated in Figure 2.
The DIA analysis applied to the entire set A uniquely determines the 
positions of 12 out of 13 microlensing events. 
In one case the recovered centroid of source's light
falls exactly in the middle between the two HST stars. Fortunately,
they have very similar colors and apparent magnitudes and so
we choose one of them at random.

\begin{figure}
\plotone{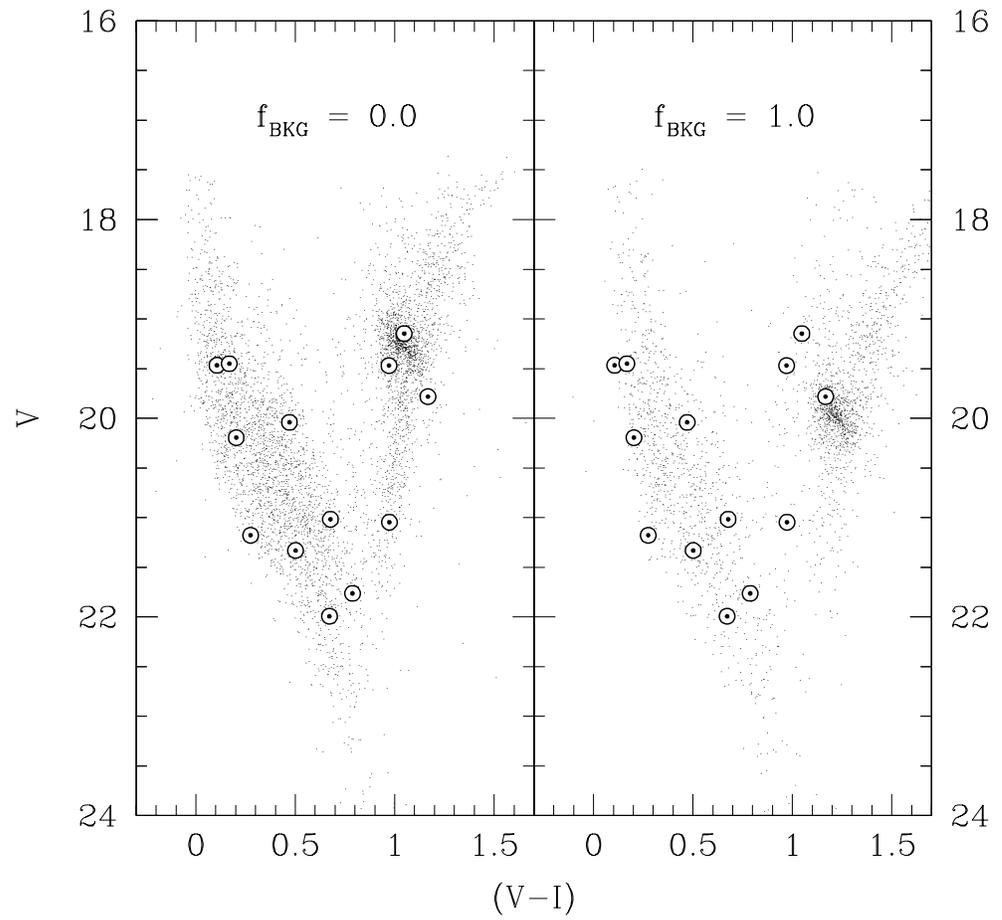}
\caption{Color-magnitude diagrams corresponding to two extreme
tested models: left one with all microlensing sources in the LMC, 
and right one
with all sources behind the LMC.}
\end{figure}

Having obtained the unblended characteristics of the source stars of
microlensing events we now have to model the source star populations
MACHO experiment is sensitive to.
We assume that our source stars can be drawn from two population:
the bar and disk of the LMC and a background population behind the 
LMC. We assume that the background population does not differ
intrinsically from the one in the LMC. The difference in the CMD
appearance comes from additional reddening (here taken to be the mean
reddening of stars in the LMC) and the shift in the distance modulus.
We start with the HST CMD and convolve it with the MACHO efficiency
for detecting stars, which goes to almost zero at $V=22$, and cuts off
a substantial fraction of the HST CMD. In this way we reproduce
the unblended LMC population observed by the MACHO experiment.
The route to mock the background population is more complicated.
Each star in the HST CMD must be first shifted by 
$\Delta (V-I) = E(V-I) = 1.38 \, E(B-V) = 0.18$ mag
and $\Delta V = A_V + \Delta_{\rm BKG}= 0.43 + \Delta_{\rm BKG}$ mag,
where $\Delta_{\rm BKG}$ is the distance modulus shift of the
background population with respect to the LMC.
Only then we modify the HST CMD by ``filtering'' the number of
stars through our efficiency curve as in the first case.
Based on these pure LMC and background populations, 
we construct a series of CMDs with a fraction $f_{BKG}$ stars
from the background and $(1-f_{BKG})$ from the main body of the LMC.
The two extreme cases of $f_{BKG}=0$ and $f_{BKG}=1$
are shown in Figure 3. Spherical symbols with central dots
are the sources of the 13 microlensing events, superposed
on the underlying population.

We perform a two-dimensional
Kolmogorov-Smirnov (KS) test to tell, which models 
are most consistent with the data.
In one dimensional case, a KS test of two samples with number of
points $N_{1}$ and $N_{2}$ returns a statistic $D$, defined to be the
maximum distance between the cumulative probability functions at any ordinate.
Associated with $D$ is a corresponding probability $P(D)$ that if two
random samples of size $N_{1}$ and $N_{2}$ are drawn from the same parent
distribution a worse value of $D$ will result.  This is equivalent to saying
that we can exclude the hypothesis that the two samples are drawn from the
same distribution at a confidence level of $1.0-P(D)$. If $N_{2} \gg
N_{1}$, then this is also equivalent to excluding at a $1.0-P(D)$ 
confidence level the hypothesis that sample 1 is drawn from sample 2.
The concept of a cumulative distribution is not defined in more than one
dimension.  However, it has been shown that a good substitute in two
dimensions is the integrated probability in each of four right-angled
quadrants surrounding a given point (Peacock 1983; Fasano \& 
Franceschini 1987). The integrated probability of each quadrant for each
distribution is the fraction of data from the distribution which lies in that
quadrant.  The two dimensional KS statistic $D$ is now taken to be the maximum
difference (ranging both over all data points and all quadrants) between the
integrated probability of distributions $N_1$ and $N_2$.  
The statistic $D$ and the corresponding $P(D)$ are subject to the same 
interpretation as in the one dimensional case (see Press et al.\ 1992
for details). 

We show the resulting $P(D)$ versus $f_{BKG}$ in Figure 4, where we
used the distance modulus shift $\Delta_{\rm BKG} = 0.3$. 
More models are considered by Nelson et al.\ (2003). 
Because the creation of the efficiency convolved CMD is a weighted random draw
from the HST CMD, the model population created in each simulation differs
slightly.  This in turn leads to small differences in the KS
statistics. The error bars indicate the scatter about the mean value
for 20 simulations. We find that the 2-D KS-test probability is
highest at a fraction of source stars behind the LMC $f_{BKG} ~ 
\sim 0.0 - 0.2$. Most of the features of considered 
CMDs are almost vertical. As a result, the significant preference for low
$f_{BKG}$ arises mostly from the reddening term (see discussion below).
In a strict statistical sense, the Kolmogorov-Smirnov test can only be 
used to reject or confirm the hypothesis that two datasets
have a common parent distribution.  
KS test results do not allow one to draw conclusions
about the relative probability, e.g., a KS test probability of 90\%
does not mean that it is
three times as likely that two distributions are identical than if
they had returned a KS test probability of 30\%. Therefore, Figure 4
should be interpreted with caution. Based on the models analyzed
by Nelson et al.\ (2003), we conclude that the general shape of the
KS probability versus $f_{BKG}$ is quite insensitive to the distance modulus
displacement.

\begin{figure}
\plotone{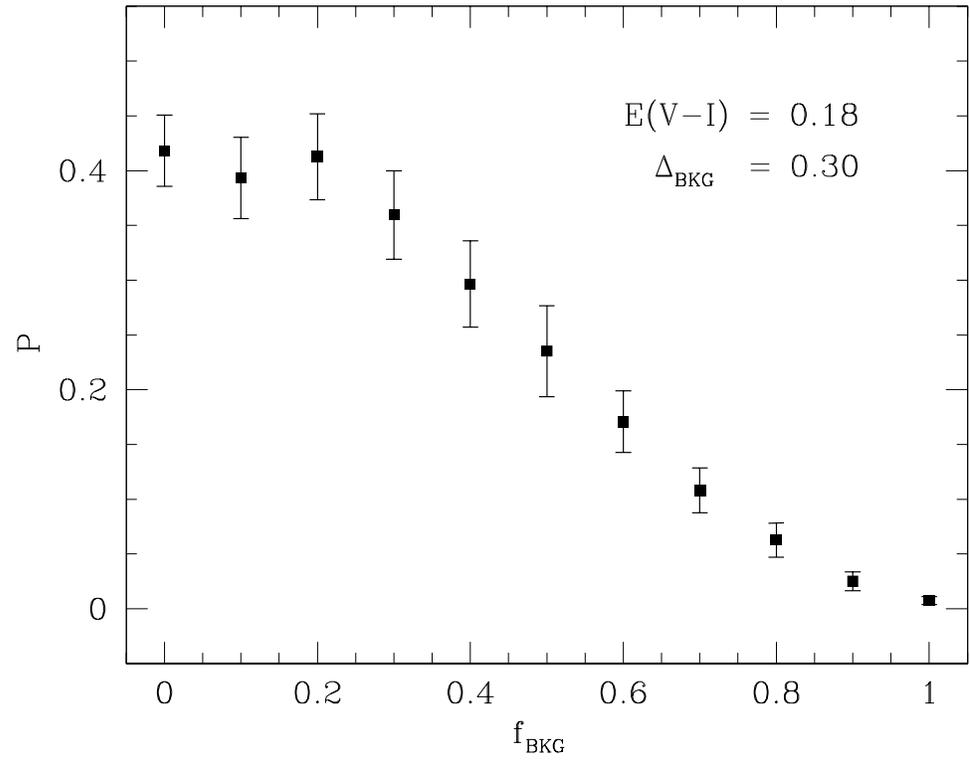}
\caption{Model consistency for different fractions of source stars
coming from the background as inferred from a 2-D Kolmogorov-Smirnov
test.}
\end{figure}

We rule out a model in which the source stars all belong to some background
population at more than a 95\% confidence level (e.g., the KS test probability
for the distance modulus displacement of $\Delta_{\rm BKG} = 0.3$ for 
$f_{BKG}>0.9$ is $\sim 0.025$).
The less extreme models of the LMC spheroid or stellar shroud
self-lensing are not excluded.
The allowed region of the $f_{BKG}$ plot ($P>10$\%) is consistent with the
expected location of the source stars in both the Milky Way lensing
and LMC disk+bar self-lensing geometries.  However, detailed modeling
of the LMC disk+bar
self-lensing suggests that it contributes at most 20\% of the observed optical
depth (Gyuk, Dalal, \& Griest 2000).  Therefore, the results of the KS test
taken together with external constraints, could suggest that the
lens population comes mainly from the Milky Way dark halo.
This statement, however, would be too strong.
Our results are not completely conclusive due to the intrinsic
problem of the method. Its serious limitation is the uncertainty
in the model of the LMC reddening. The significance of our test will decrease
substantially if the average reddening of the LMC is significantly
lower or the reddening is very patchy.

\section{Direct identification of a microlens}

\begin{figure}
\plotone{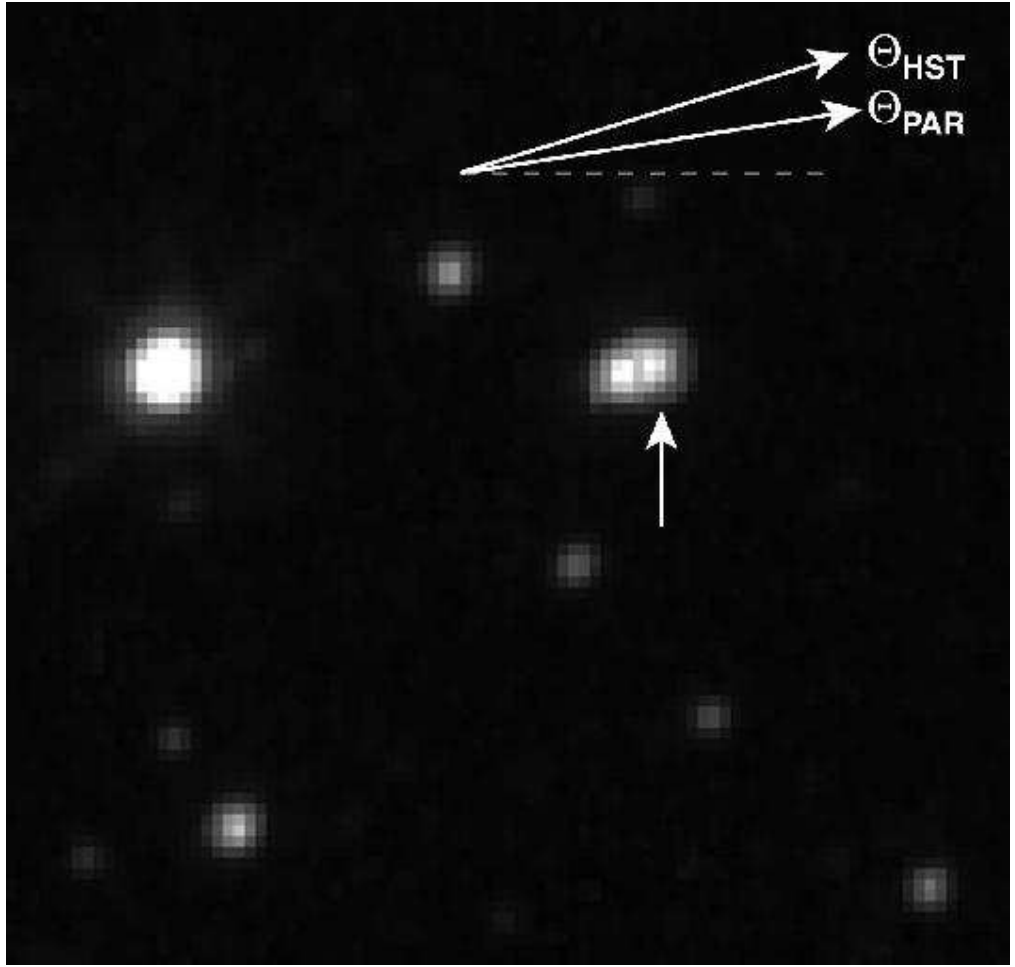}
\caption{LMC-5 microlensing event: image showing the lens (arrow) 
and the source next to it. The arrows in the upper right corner
indicate the direction of the lens motion obtained from the parallax
fit and HST imaging.}
\end{figure}

The HST Wide Field Planetary Camera (PC) observations described in the previous
section included imaging of the LMC-5 event (Alcock et al. 2001c). 
Image of this event in HST V, R, and I filters 
was taken on May 13, 1999, 6.3 years after the peak on February 5, 1993.
The system, composed of two objects displaced by 0.134'', 
was resolved with PC pixels of 0.046'' (Figure 5). One of these objects
was shown to be a main sequence source stars in the LMC with
$V = 21.02 \pm 0.06$, $(V-R) = 0.30 \pm 0.09$, and $(V-I) = 0.68 \pm
0.10$. Its neighbor, marked with an arrow, is a very red, faint
object with $V = 22.67 \pm 0.10$, $(V-R) = 1.60 \pm 0.12$, and 
$(V-I) = 3.18 \pm 0.11$. 
A chance superposition of such a red
object on the LMC source star is of order of $10^{-4}$.
Therefore, it is tempting to assume that the red object is the lens.
The light curve of event LMC-5 showed very large maximum flux 
amplification $A_{\rm max}$,
which corresponds to very small impact parameter. As a result, we
expect that at the peak, the angular distance between the lens and 
the source was negligible compared to the measured separation.
Indeed, for large amplification events the angular distance between
the lens and the source is at the peak equal to
$4.036 \cdot 10^{-4}$''  
$\! \times A_{\rm max}^{-1} \sqrt{1/x-1} \sqrt{m_{\rm lens}/M_{\odot}}$,
where $x = D_l/D_s$. For LMC-5, the term 
$A_{\rm max}^{-1} \sqrt{1/x-1} \sqrt{m_{\rm lens}/M_{\odot}}$
is very likely to be $ < 1$, and, as a result, the lens-source
separation at the peak was of order of or smaller than $10^{-3}$ of
the current separation.
Consequently, the current angular separation is an excellent measure of the
relative proper motion between the source and the lens, 
$\mu_{\rm rel} = 0.0214 \pm 0.0007$ arcsec/yr with the direction,
$\Theta_{\rm rel} = -92^{\circ}$, given by the line connecting 
the light centroids.

\begin{figure}
\plotone{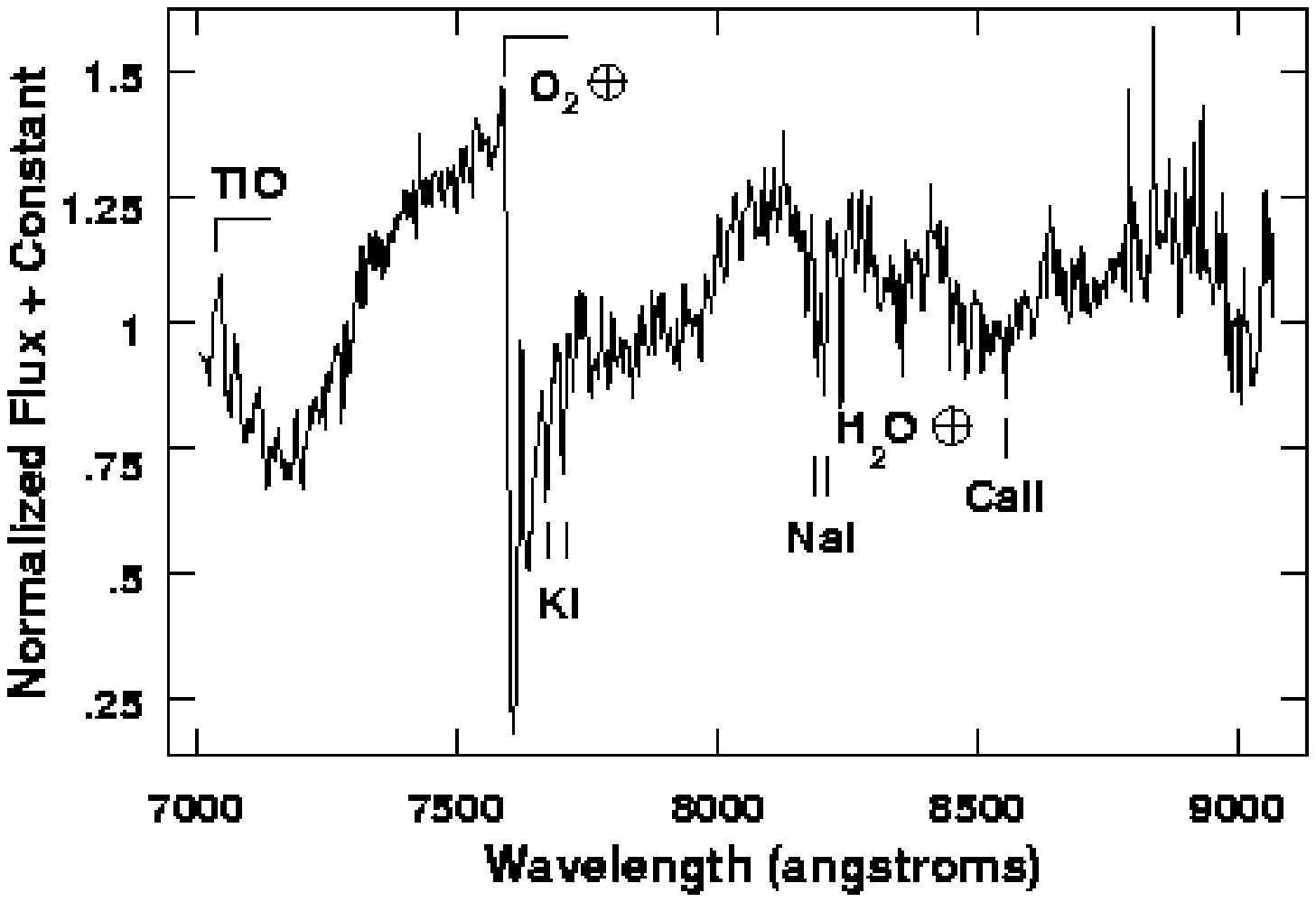}
\caption{Blended spectrum of the source and the lens. The red part
is dominated by the lens.}
\end{figure}

Interestingly, LMC-5 has a significant detection of the microlensing
parallax effect. We obtained the projected transverse velocity
$\tilde{v} = 18$ km/s, the direction of lens proper motion, 
$\Theta_{\rm par} = -97^{\circ}$, and duration $t_E = 22.5$ days.
The agreement between $\Theta_{\rm par}$ and $\Theta_{\rm rel}$ suggests that
the red object is indeed the lens for the microlensing event LMC-5. 
Equation (4)
%(\ref{mass}) 
can therefore be used to estimate
the mass of the lens $m_{\rm lens} = 0.036^{+0.009}_{-0.004} M_{\odot}$,
which argues for a sub-stellar object at a very high significance.
The errors, however, are not Gaussian and the $3\sigma$ limit
on the mass based on the parallax fit is $m_{\rm lens} \leq 0.097
M_{\odot}$.
In natural units, the relative parallax of the event
is $\pi_{\rm rel} = \mu_{\rm rel}/\tilde{v}$. For a nearby lens and
a source in the LMC, $\pi_{\rm rel} \approx \pi_{\rm lens}$.
This allows us to derive the distance to the lens:
$d_{\rm lens} = 200^{+40}_{-30}$ pc.
Assuming no attenuation of light due to Galactic extinction, we obtain
the absolute magnitude of 
$M^{200 {\rm pc}}_V = 22.67 - 5.0 \log (200 {\rm pc}/10 {\rm pc}) = 16.16$.

We took the spectrum
of the LMC-5 microlensing system with the European Southern
Observatory Very Large Telescope on February 2,
2001 .
The separation of the lens-source system at this time was about
0.2'', which was unresolved and resulted in obtaining a composite
spectrum (Figure 6). This spectrum becomes more and more dominated by the lens
flux as we move redwards of 7000 $\!$\AA. The presence of KI, NaI,
the absence of CsI, RbI, and the TiO band at 7100 $\!$\AA $\,$ coupled with
the absence of the VO band at 7450 $\!$\AA $\,$ lead to conclusion that
the lens is of spectral type M4-5V. Our HST colors are consistent
with this spectral classification (Bessell 1991).
According to Cassisi et al.\ (2000) this spectral type implies
$m_{\rm lens} \sim 0.095-0.013 M_{\odot}$, which is in $3\sigma$ 
disagreement with the parallax result.

If the lensing star is an M-dwarf, then it should follow
the empirical relation
\begin{equation} 
M_V  = 2.89 + 3.37(V-I)_0 \label{mvreid}
\end{equation}
derived by Reid
(1991). Assuming zero reddening, we obtain
$M_V = 13.61 \pm 0.55$ and $d_{\rm lens} = 650 \pm 190$ pc, again in
disagreement with parallax result.

Can the reddening be the reason of this discrepancy?
In general:
\begin{equation}
M_V = V -5\log \frac{d}{10 {\rm pc}} - A_V \label{mvgeneral}
\end{equation}
where $d$ is the distance and $A_V$ is the visual extinction, which
can be written as $A_V = R_{VI} E(V-I)$. Color excess
is defined as $E(V-I) \equiv (V-I) - (V-I)_0$.
Combining equations (5)
%(\ref{mvreid}) 
and (6) %(\ref{mvgeneral}) 
for our gravitational lens yields:
\begin{equation}
d_{\rm lens} ({\rm pc}) = 10^{0.2 (5.0 +
V-2.89-3.37(V-I)+(3.37-2.50)E(V-I))}
= 10^{2.815 + 0.174 \, E(V-I)}, \label{dlens}
\end{equation}
where we assumed a standard value for the selective extinction
coefficient $R_{VI} = 2.50$. Equation (7) %(\ref{dlens}) 
shows that a non-zero extinction would only increase the disagreement between
the parallax fit and the spectroscopically motivated estimate of
the lens distance (Figure 7). If the lens is not at the very
unusual line of sight, then $E(V-I) \sim 0.1$ and correction due
to reddening is small, well within the errors of our estimate
based on the assumption of zero extinction.

A heliocentric radial velocity of the lens computed using potassium
and sodium lines is equal to $v_{\rm rad} = 49 \pm 10$ km/s.
Assuming a distance to the lens and the proper motion inferred from
the HST images we solve for the space velocity.
At 200 pc, the velocity of the lens relative to the local standard
of rest is $U = 27$ km/s, $V=-43$ km/s, and $W=-4$ km/s, 
toward the Galactic center, along rotation, and up out of the plane,
respectively. Therefore, the lens has disk-like kinematics.
Motivated by the surprisingly faint absolute magnitude
implied by the solution based on the parallax fit, 
Alves \& Cook (2001) considered a possibility that
the LMC-5 lens is a halo subdwarf. However, since
metal-poor stars with disk-like kinematics
are very rare, there is a strong indication that the lens 
belongs to the Galactic disk.

\begin{figure}
\plotone{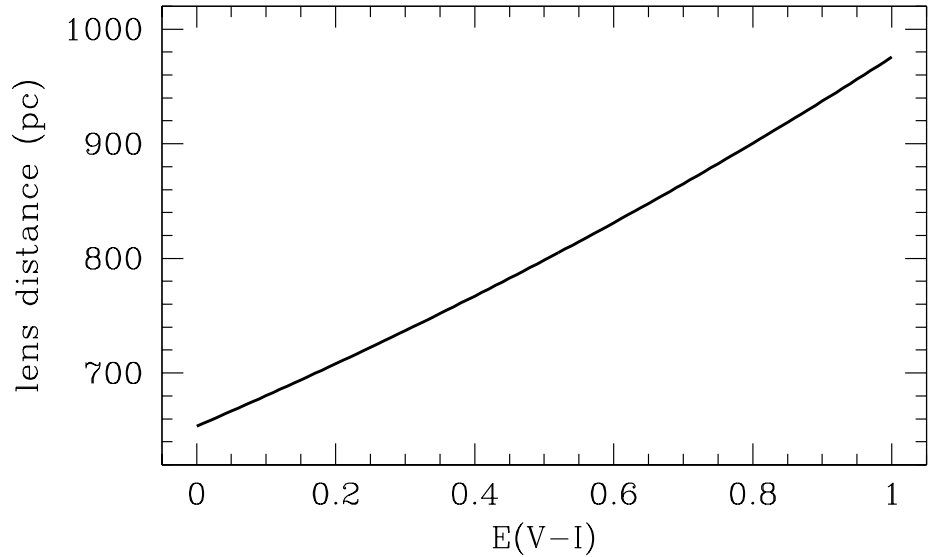}
\caption{The distance of the lens as a function of its reddening under
the assumption that the lens is an M dwarf that follows the $M_V$ --
$(V-I)_0$ relation of Reid (1991). The non-zero
reddening with a standard extinction curve requires
larger $d_{\rm lens}$ and so increases the
discrepancy between the parallax fit results and photometric
constraints that follow from the spectral classification.}
\end{figure}

The inconsistencies between solutions based on the HST-based proper motion
coupled with the MACHO parallax fit on one side and HST photometry
coupled with the VLT spectrum on the other might have arisen from 
difficulty of measuring
parallax effect for such a short event. 
This conflict will be resolved with observations we took on
HST Advanced Camera for Surveys. Thanks to a very small pixel size of 
0.027'', two epochs of data will allow us to verify the proper motion of
the lens and measure the parallactic motion. 
Despite the described inconsistencies we believe that the red object
detected in our images is the lens for the LMC-5 event and
that it is a normal low-mass disk star.

\section{Microlensing optical depth toward the Galactic bulge}

After almost a decade of surveys, 
the constraints from the Milky Way microlensing observations 
imposed on the 
dark matter were so tight (Binney \& Evans 2001) that they challenged
the standard cuspy dark halo profiles (Navarro, Frenk, \& White 1997;
Moore et al.\ 1999).
On the other hand, these cuspy profiles are very likely a generic feature of 
the cold dark matter cosmology.
Below we briefly describe the measurements of the optical depth toward
the Galactic bulge that led to this situation and proceeded the study
described here.
In brief, the studies discussed below yield the microlensing optical depth
toward the Galactic bulge in the range $\tau_{\rm bulge} = $3$-$4 $ \times 10^{-6}$, 2$-$3
times higher than implied by other observational constraints and
theoretical models.
Udalski et al.\ (1994) found 9 events in the first two year of the Optical
Gravitational Lensing Experiment (OGLE) data. They set the lower limit
on the optical depth to the Galactic bulge at $\tau_{\rm bulge} = (3.3 \pm 1.2) \times
10^{-6}$. The uncertainties in this study were related to the detection
efficiency analysis as well as small number statistics.
Alcock et al.\ (1997a) described a set of 45 bulge events.
They computed the so called sampling efficiencies, which are a
good approximation only for bright events.
Therefore, unbiased analysis could have been done only for 13 clump
giants, which resulted in large uncertainties of the optical depth
($\tau_{\rm bulge} = 3.9^{+1.8}_{-1.2} \times 10^{-6}$).
Udalski et al.\ (2000) presented a catalog of over 200 microlensing events
from the last 3 seasons of the OGLE-II bulge observations. 
Wo\'{z}niak et al.\ (2001) released another catalog of about 500
events selected using image subtraction technique.
Unfortunately, 
no efficiency analysis has been done for these two samples of OGLE
events so the information 
that can be extracted from them is very limited.
Alcock et al.\ (2000a) performed Difference Image Analysis of three
seasons of bulge data in 8 frequently sampled MACHO fields and found 99
events. They determined $\tau_{\rm bulge} = 3.23^{+0.52}_{-0.50}
\times 10^{-6}$ at $(l,b) = (2\fdg 68, -3 \fdg 35)$.
This was a major development in bulge microlensing.
The DIA technique resulted in a substantial improvement in photometry, so
this analysis was less vulnerable to uncertainties in the parameter 
determination.
However, there are two potential problems with this analysis. First,
the detection efficiency estimate suffers from the fact
that a deep HST luminosity function was available for only one field.
Second, the
lensed sources are not guaranteed to be in the Galactic bulge. The
measured optical depth is, therefore, converted to the optical depth
toward the bulge using a fudge factor. Popowski (2002) showed that
the reasonable modification of this factor alone can lower the
estimate of the optical depth by almost 20\%. To cure this
uncomfortable situation we would like to select the event that
have sources in the bulge. Events with clump giants as sources are
excellent candidates, since clump giants are among the brightest
and most numerous stars in the bulge. Being bright, clump sources
have an additional nice feature -- they are almost unaffected by blending.

\begin{figure}
\plotone{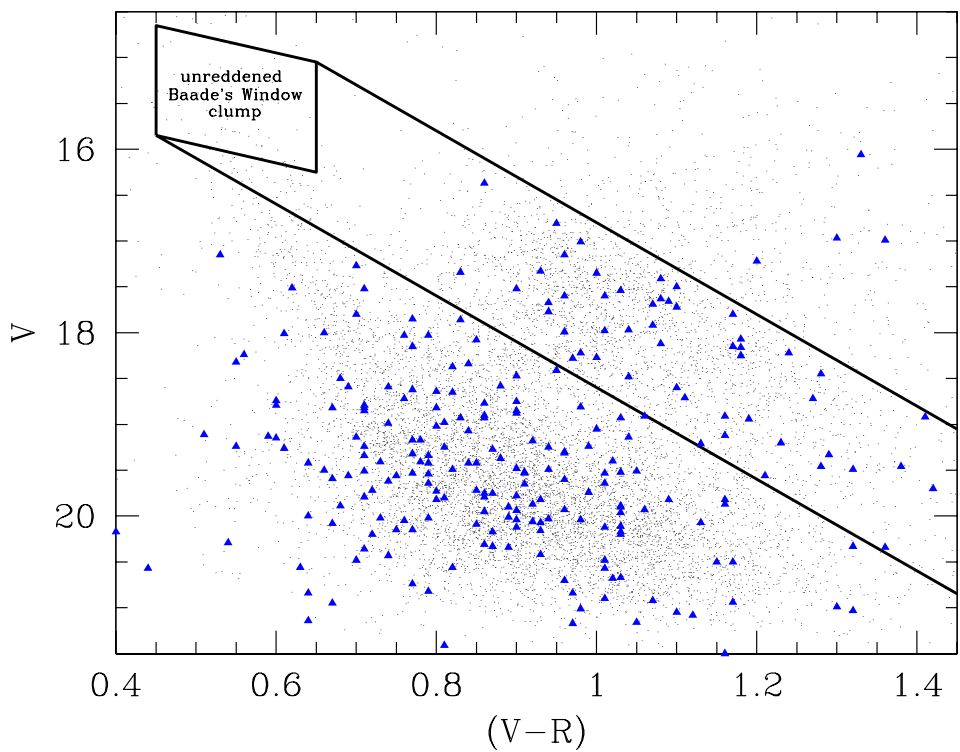}
\caption{The locus of clump giants in a color-magnitude diagram can
be obtained by shifting the locus of unreddened clump population at
the Galactic bulge distance along the reddening line. Here we show the
clump region obtained under the assumption that the coefficient of
selective extinction $A_V/E(V-R) = 5.0$.}
\end{figure}

Blending is a major problem in any analysis of the 
microlensing data involving point spread function photometry.
The bulge fields are crowded, so that
the objects observed at a certain atmospheric seeing are blends of several 
stars.
At the same time, typically only one star is lensed.
This complicates a determination of the events' parameters
and the analysis of the detection efficiency of microlensing
events.
If the sources are bright one can avoid these problems. First, a 
determination of parameters of the actual microlensing events becomes
straightforward. Second, it is sufficient to estimate detection efficiency
based on the sampling of the light curve alone. This eliminates the need
of obtaining deep luminosity functions across the bulge fields. 
Therefore, in this discussion we concentrate
on the events where the lensed stars are clump giants
(Popowski et al.\ 2001a,b). We are using the results of the analysis
performed on five seasons of data (1993--1997) in 77 fields.

\begin{figure}
\plotone{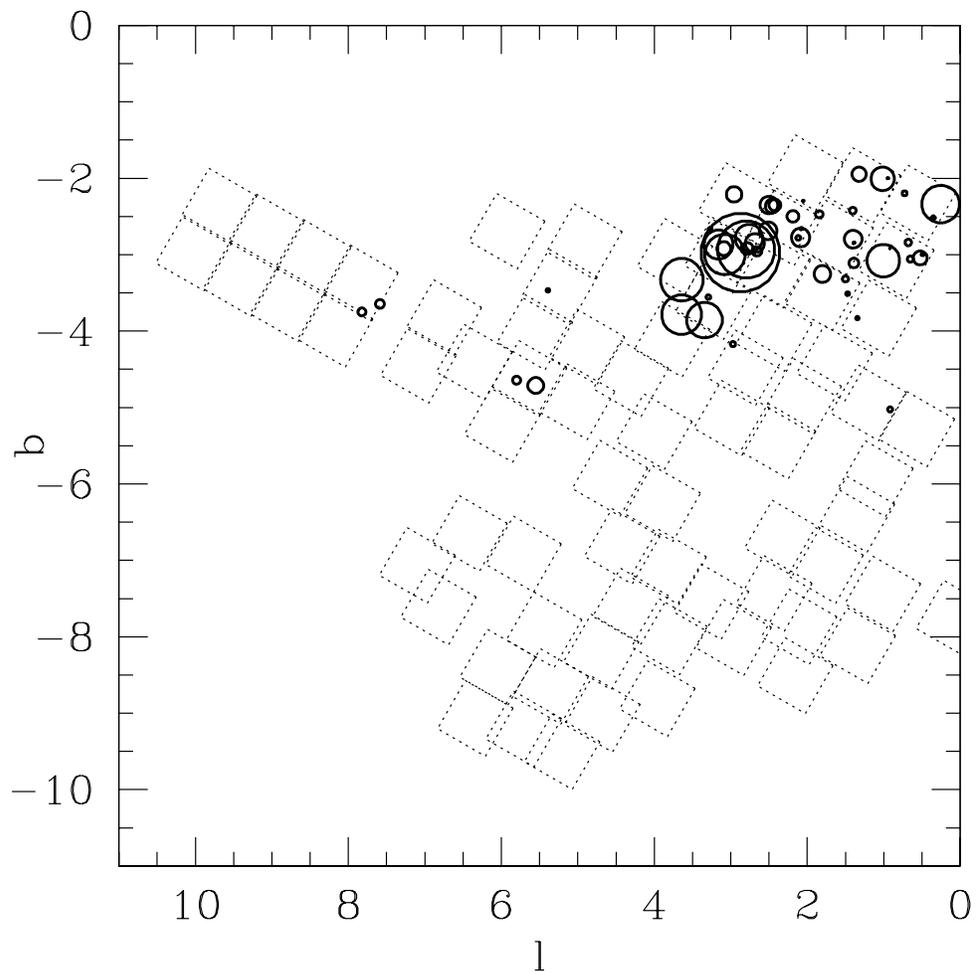}
\caption{The location of the analyzed MACHO fields (square regions) with the
location of the clump events (circles). The size of each circle is proportional
to the duration of the event. Note the concentration of long events
in field 104 at $(l,b) = (3 \fdg 1, -3 \fdg 0)$.}
\end{figure}

The events with clump giants as sources have been selected from
the sample of all events. The procedure that leads to a selection
of microlensing events of general type consists of several steps.
First, all the recognized objects in all fields are tested for any 
form of variability.
Second, a microlensing light curve is fitted to all stars showing any variation
and the objects that meet very loose selection criteria (cuts) enter the next 
phase. Here, this selection returns almost 43000 candidates.
These candidates undergo more scrutiny and are subject to more
stringent cuts, most of which test for a signal-to-noise of
the different parts of the light curve. Here, this last procedure narrows
a list of candidate events to $\sim 280$.
The question, which of those sources are clump giants, is investigated
through the analysis of the global properties of the color-magnitude
diagram in the Galactic bulge.
Using the accurately measured extinction towards Baade's Window 
(Stanek 1996 with zero point corrected according to Gould, Popowski,
\& Terndrup 1998 and 
Alcock et al.\ 1998) allows us to locate bulge clump giants on the 
dereddened color-magnitude diagram. 
Such diagram can be then used to predict the positions of clump
giants on the color-apparent magnitude diagram for fields with different
extinction.
Based on Baade's Window data we conclude that unreddened clump giants
are present in the color range $(V-R)_0 \in (0.45,0.65)$, where
they concentrate along a line $V_0 = 14.35 + 2.0 \, (V-R)_0$.
We assume that the actual clump giants scatter in $V_0$-mag around these 
central values, but
by not more than 0.6 mag toward both fainter and brighter $V_0$.
This defines the parallelogram-shaped box in the upper left corner
of Figure 8. With the assumption that the clump populations in the whole bulge
have the same properties as the one in the Baade's Window,
the parallelogram described above
can be shifted by the reddening vector
to mark the expected locations of clump giants in different fields.
The solid lines are the boundaries of the region where one could find
the clump giants in fields with different extinctions.
There are a few more $V$-mag and $(V-R)$-color cuts that determine the
final shape of the clump region. 
Several assumptions that went into creating this region should be
carefully reviewed. 
For example, 
the assumed spread in $V$ magnitudes can be either bigger or smaller
or asymmetric around the central value, clump giants in different fields may 
have different characteristics etc.
In retrospect, we conclude that the effect of the uncertainty in
the selective extinction coefficient $R_{VR}$ is probably the hardest
to deal with\footnote{All the selection-related problems are treated 
in depth in the analysis of 8 years of microlensing data.}.
The clump region from Figure 8 contains 52 unique
clump events from the five bulge seasons discussed here. 
Their location in Galactic coordinates is shown in Figure 9,
superposed on the outlines of the analyzed MACHO fields.
The size of each circle is proportional to the duration of the event.

Based on this sample of events we estimate the bulge optical depth
from the following formula:
\begin{equation}
\tau = \frac{\pi}{2NT} \sum_{\small\rm all \; events} \frac{t_E}{\epsilon(t_E)}, \label{optdepth}
\end{equation}
where $N$ is the number of observed stars (here about 2.1 million clump 
giants), $T$ is the total exposure (here about 2000 days) and 
$\epsilon(t_E)$ is an efficiency for detecting an event with a given $t_E$.
The sampling efficiencies were obtained with the pipeline that has been
previously applied to the LMC data (for a description see 
Alcock et al.\ 2001d). 
In brief, artificial light curves with different parameters have been
added to 1\% of all clump giants in the 77 considered fields and the 
analysis used
to select real events was applied to this set. For a given duration of the
artificial event, the efficiency was computed as a number of recovered events
divided by a number of input events.
The efficiencies used in this analysis are global efficiencies averaged over
clump giants in all 77 fields. The optical depth is reported at the
central position that is an average of positions of 1\% of the analyzed clump 
giants. We obtain:
\begin{equation}
\tau_{\rm bulge} = (2.0 \pm 0.4) \times 10^{-6}\;\;\;\;\; {\rm at} \;\;\;\;\; (l,b) = (3 \fdg 9, -3 \fdg 8). \label{optDepth}
\end{equation}
with the error computed according to the formula
given by Han \& Gould (1995).
In Figure 10 we plot the spatial distribution of the optical depth. 
The variation of the optical depth is dominated by the Poisson noise.
The gradient of the optical depth is stronger in $b$ than in $l$.
About 40\% of the optical depth is in the long events with
$t_E > 50$ days. Such a population of long events is not easily
explained by the standard models of Galactic structure and kinematics.
Field 104 at $(l,b) = (3 \hbox{$.\!\!^\circ$} 1, -3
\hbox{$.\!\!^\circ$} 0)$
has the largest number of clump events (10) and the optical
depth of $(1.4 \pm 0.5) \times 10^{-5}$.

\begin{figure}
\plotone{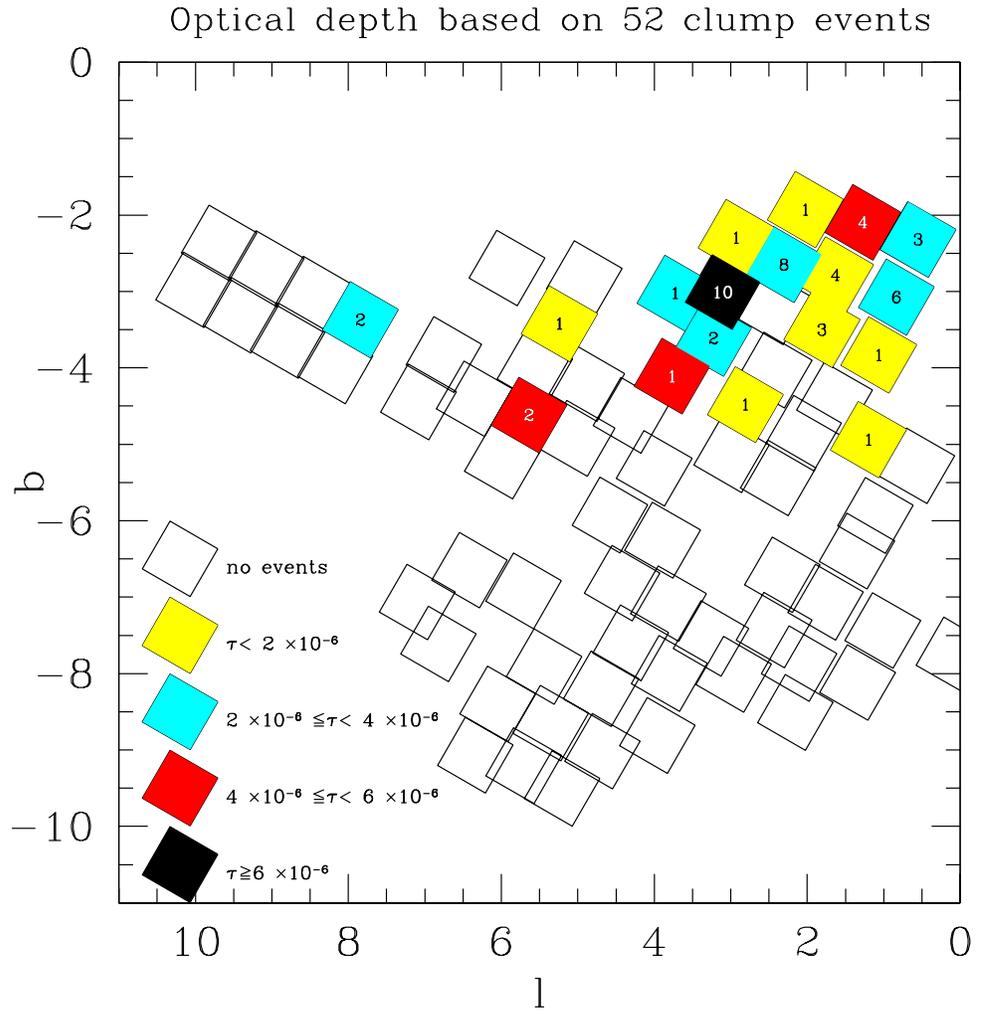}
\caption{Field-by-field optical depth. The shades correspond
to different ranges in the optical depth. The numbers of events
are given in the center of each field. White fields with no numbers
have no detected events. The errors in the optical depth are dominated
by the Poisson noise. Note the high optical depth in field
104 at $(l,b) = (3 \hbox{$.\!\!^\circ$} 1, -3
\hbox{$.\!\!^\circ$} 0)$.}
\end{figure}

There is a high concentration of long-duration events in this field (5 out
of 10 events longer than 50 days are in 104, including the longest 2).
We investigate how statistically significant is this concentration. 
The analysis of event durations uncorrected for 
efficiencies provides a lower limit on the difference between the
frequently-sampled field 104 and all the remaining clump giant fields.
We use the Wilcoxon's number-of-element-inversions statistic 
(see Popowski et al.\ 2001a for a description). 
The Wilcoxon's statistic is equal to 320,
whereas the expected number is 210 with an error of about 43. Therefore
the events in 104 differ (are longer) by $2.55 \sigma$ from the other fields.
That is, the probability that events in 104 and other fields originate from
the same parent population is of order of 0.011.
Popowski's (2002) analysis of the DIA events from 
Alcock et al.\ (2000a) confirms this duration difference.
We conclude that field 104 is anomalous both in terms of the optical
depth and duration distribution. Both effects can be explained 
simultaneously by the concentration of mass along this line of sight.

The currently available Galactic models are smooth on the
scales of a single MACHO field. Therefore, they cannot account for
such a localized anomalous behavior as in field 104, and it seems justified to
remove this field as an outlier.
With this modification, the optical depth decreases to 
$\tau_{\rm bulge} = (1.4 \pm 0.3) \times 10^{-6}$, which is $2/3$ of its original
value, and becomes fully consistent with infrared-based models of the 
Galactic bar (Bissantz \& Gerhard 2002, Evans \& Belokurov 2002).
This lower optical depth is supported by the very
recent determination from the EROS group by Afonso et al.\ (2003),
who report $\tau_{\rm bulge} = (0.94 \pm 0.29) \times 10^{-6}$ 
based on 16 clump events in fields centered on 
$(l,b) = (2 \hbox{$.\!\!^\circ$} 5, -4\hbox{$.\!\!^\circ$} 0)$.
Surprisingly, another recent determination of the optical depth
by the MOA collaboration (Sumi et al.\ 2002), based on 28 events
recovered with DIA technique, gives
the high optical depth of 
$\tau_{\rm bulge} = 3.36_{-0.81}^{+1.11} \times 10^{-6}$. This value compares
very well with the DIA result of Alcock et al.\ (2000a).
In conclusion, there is no longer a general discrepancy between the 
microlensing observations and Galactic models. Instead, there is a puzzling
gap between the optical depth measurements based on clump events
and the samples from the Difference
Image Analysis technique.

\section{Stellar remnants as gravitational microlenses}

Equations (1) and (2) show that durations of microlensing events could in
principle be a good probe of lens masses. 
However, event's duration depends also on
the kinematics of populations involved
in the lensing process. As nicely showed by Gould (2000; Figure 1),
the duration distribution
is very wide even for a population of lenses with a single mass.
Therefore, despite the fact that $\sim 20$ \% of events may be due
to remnants in the form of white dwarfs, neutron stars and black
holes, one cannot use just durations to identify this population.
Events showing parallax effect (equation 4) %\ref{mass})
are in high demand as they provide very stringent constraints on the
lens masses.

\begin{figure}
\plottwo{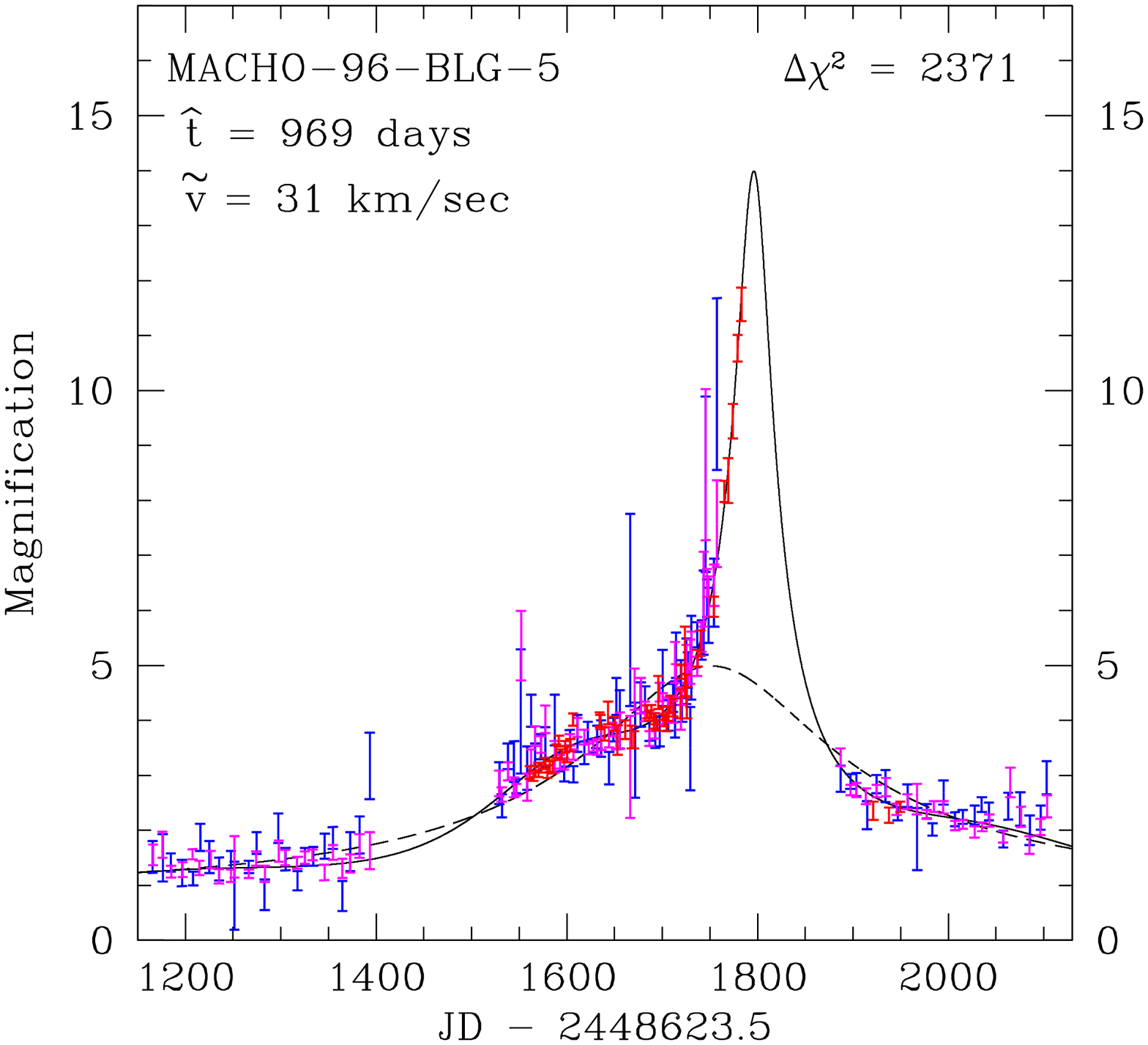}{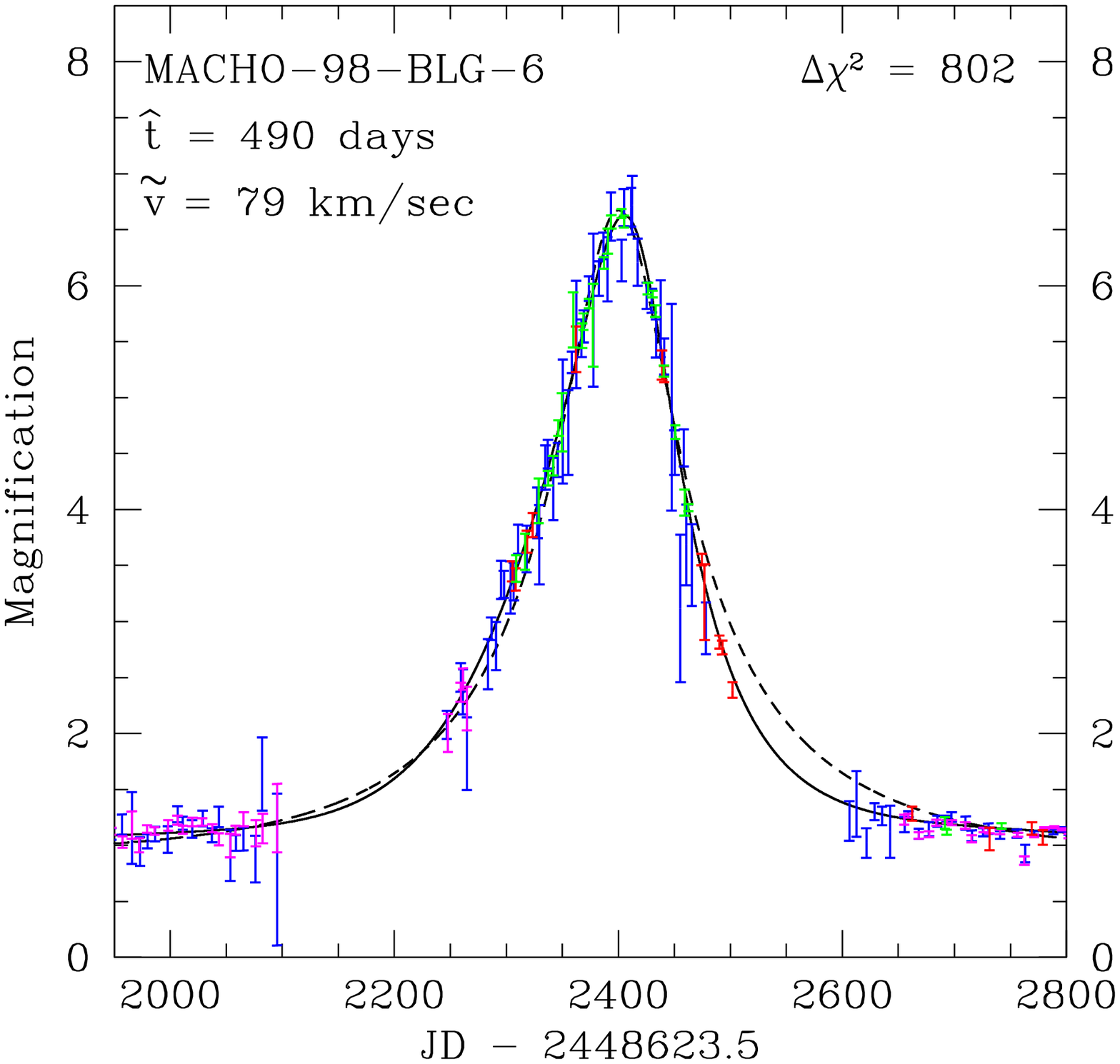}
\caption{Light curves of two microlensing events with likely black
hole lenses. The data come from the MACHO project and several sources
of follow-up observations. 
The improvement of the parallax fit (solid curve) over the standard
fit (dashed curve) is easily visible.}
\end{figure}

We present the analysis of the longest timescale microlensing
events with $t_E > 70$ days (Bennett et al.\ 2002).
We used the MACHO experiment data described in \S 1 as well as
the follow up data from the MACHO/GMAN observations on the CTIO
0.9 m telescope (Becker 2000) and from the MPS Collaboration 
on the Mount Stromlo 1.9 m telescope (Rhie et al.\ 1999).
We performed parallax fit to over 300 events
detected by MACHO toward the Galactic bulge. We considered parallax
signal to be reliable when the difference in the goodness of fit
between the standard microlensing curve and the parallax fit 
was $\Delta \chi^2 \geq 200$.
This selection criterion returned 6 events:
104-C, 96-BLG-5, 96-BLG-12, 98-BLG-6, 99-BLG-1, and 99-BLG-8.
The light curves and microlensing fits for the two microlensing events
with likely black hole lenses, 96-BLG-5 and 98-BLG-6, are presented
in Figure 11.

The CMD analysis of the events indicates that MACHO-104-C and
MACHO-96-BLG-12 source stars are bulge clump giants.
Source star for MACHO-99-BLG-8 is likely a bulge giant.
MACHO-96-BLG-5 is likely a blended main sequence star.
The CMD locations of MACHO-98-BLG-6 and MACHO-99-BLG-1 source
stars suggest that they can be either bulge subgiant stars or
red clump giants in the Sgr dwarf galaxy. However, their radial velocities
of $v_{\rm rad} = -65 \pm 2$ km/s and $v_{\rm rad} = 64 \pm 2$ km/s,
respectively are not consistent with the radial
velocity of Sgr dwarf $v_{\rm rad}^{\rm SGR} = 140 \pm 10$ km/s  
(Ibata et al.\ 1997). Therefore, we concluded that they must have
bulge sources.

\begin{figure}
\plottwo{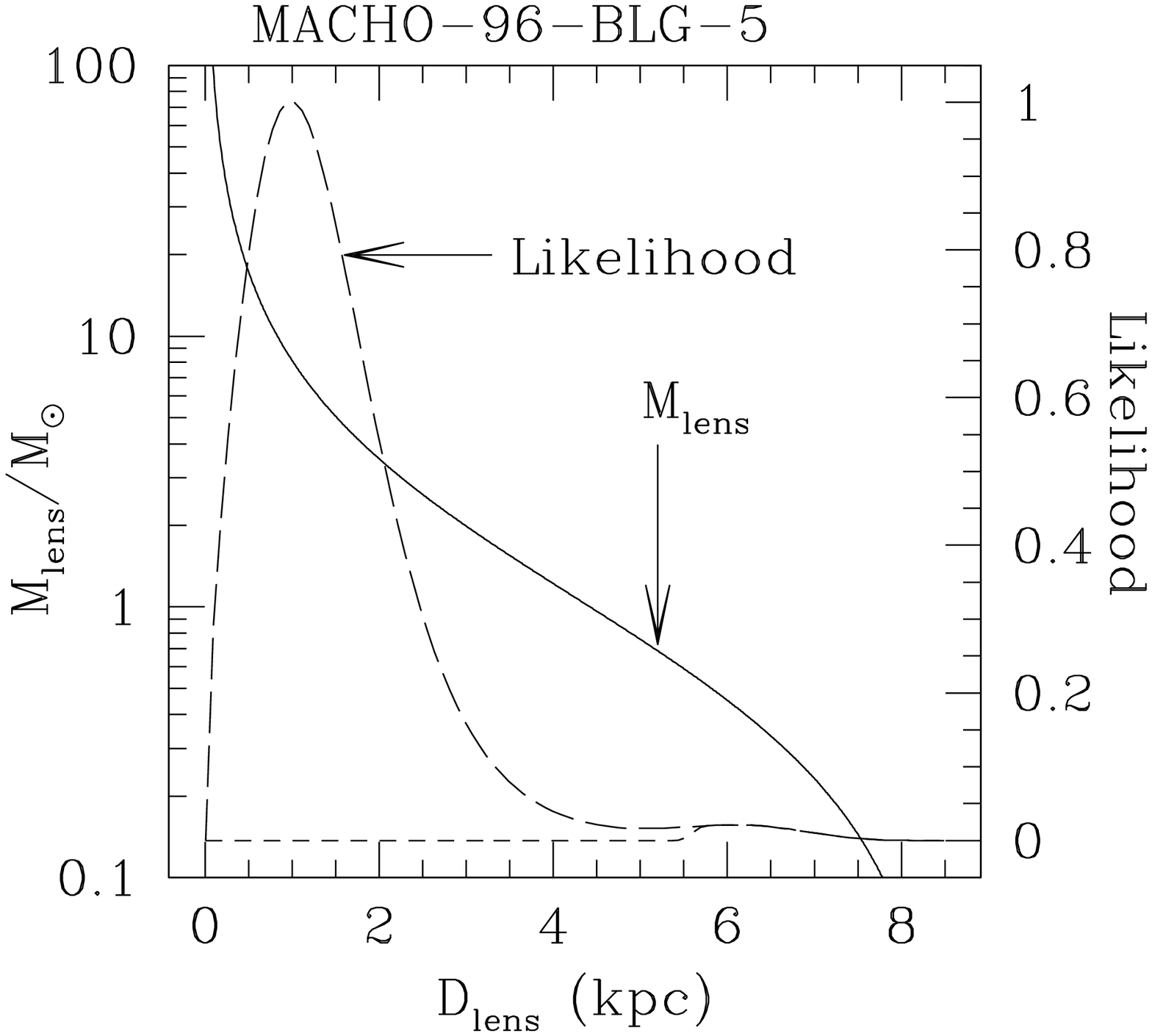}{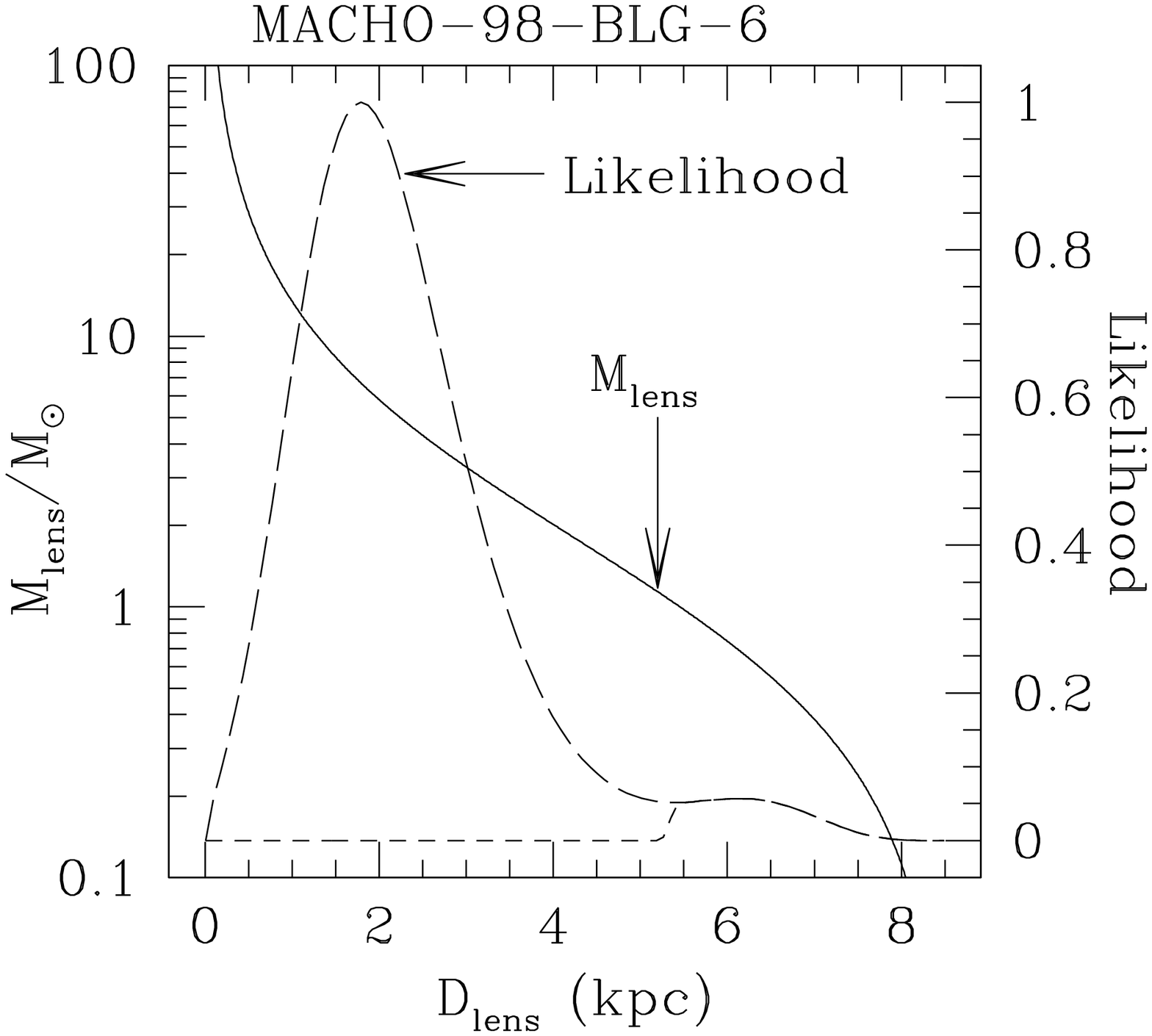}
\caption{Mass versus distance relations for two microlensing events
with likely black hole lenses (solid lines). The likelihood functions 
computed using
a standard Galactic model are shown as long-dashed lines.
The short-dashed line gives a lower limits on the distance to the lens
assuming that the lens is a main sequence star. Such a case
is particularly unlikely for event 96-BLG-5.}
\end{figure}

The 6 parallax events have $\tilde{v}$ in the range $30-80$ km/s.
If we were to assume that the bulge and disk velocity dispersions
are negligible in comparison with the Galactic rotation of about
200 km/s, then, for the disk lenses, equation (3) would
imply that $D_l = D_s \tilde{v}/(200 {\rm \,km/s} + \tilde{v})$.
For the sources in the bulge at $D_s \approx 8$ kpc and our measured
$\tilde{v}$ this would result in the distances of the lenses between
about 1 kpc and 2.3 kpc. In reality, the bulge and disk velocity dispersions
modify this result to some extent, but this example shows how
we can obtain constraints on the lens distance (and consequently
mass) from kinematics of Galactic populations.
Here we assume that the stellar remnants have
the same kinematics and density distributions as the observed
stellar populations.
To construct likelihood functions of the lens distance given its
$\tilde{v}$, we assume for the density profiles a standard double 
exponential disk and barred bulge of Han \& Gould (1996).
Our model disk has an isotropic velocity dispersion of 30 km/s,
and a flat rotation curve of 200 km/s. The bulge is
non-rotating and has an isotropic velocity dispersion of 80 km/s.
The resulting likelihood curves are shown in Figure 12 as dotted
lines for the two
parallax events with the highest predicted lens masses.
The solid line is a graphic representation of equation (4) % (\ref{mass})
under the assumption that the sources are in the
bulge at $D_s = 8$ kpc.
Following the Bayesian method and assuming a uniform prior, we
may interpret our likelihood function as a probability distribution
of a certain lens mass, which allows us to determine the uncertainty
in mass estimation.

We may obtain an additional constraint on the lens distances and
masses if we assume that
they are main sequence stars. This is achieved by finding the
upper limit on the lens brightness. To this end, we use the blending
fraction derived from the parallax fit. For our two best candidates:
the main sequence lens is extremely unlikely for MACHO-96-BLG-5;
for MACHO-98-BLG-6, the main sequence lens is not excluded and
its most likely distance is about 6.1 kpc.
These additional constraints are depicted in Figure 12 by short-dashed
lines.

\begin{figure}
\plotone{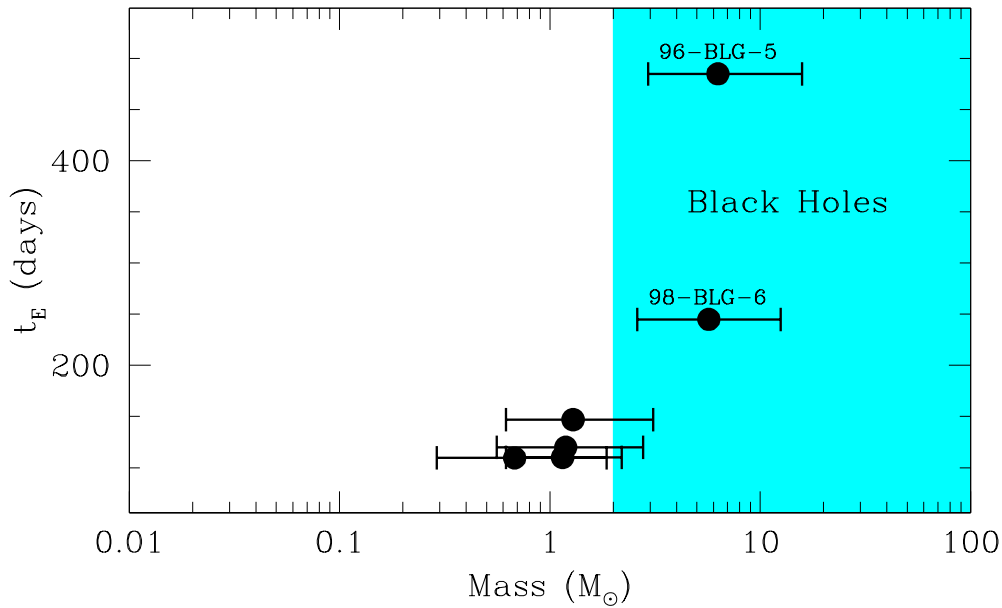}
\caption{Event duration versus the lens mass for the 6
events with strong parallax signal. One sigma error bars are shown.}
\end{figure}

In Figure 13 we present the mass estimates for 6 bulge events
with strong detections of the parallax signal. The region of likely
black holes, i.e., the lenses with masses larger than $2 M_{\odot}$
(Akmal, Pandharipande, \& Ravenhall 1998)
is shaded. The 95\% lower limits on the masses of two candidates
that fall into this region are: 1.64 $M_{\odot}$ for MACHO-96-BLG-5
and 0.94 $M_{\odot}$ for MACHO-98-BLG-6. Therefore, in principle both of
them could be neutron stars and for MACHO-98-BLG-6 even the main
sequence star is a possibility. But we note that the measured 
neutron stars masses are close to the Chandrasekhar mass, 
$M_{NS} = 1.35 \pm 0.04 M_{\odot}$ (Thorsett \& Chakrabarty 1999).
Such masses are excluded for MACHO-96-BLG-5 at more than 95\%
confidence and for MACHO-98-BLG-6 at more than 90\% confidence,
which makes them excellent black hole candidates.
We obtain
similar conclusions when we generalize our likelihood
function to include the lens mass function prior as suggested by Agol
et al.\ (2002), but the results are very sensitive to the assumed 
mass function.

Together with MACHO-99-BLG-22/OGLE-1999-BUL-32 (Mao et al.\ 2002),
the lenses of these events are the first candidates for black holes
detected without observations of matter bound to the black hole.
These 2 (3) events come from the considered sample of around 300
events, thus constituting around 1\% of all events. However, they have
very long durations, and contribute substantially more to the
total amount of mass
along the line of sight. If all three events are causes by black
hole lenses, then the stellar mass fraction of black holes may be
as high as 10\%. This in turn would suggest that most of the black
holes are not observed as X-ray binary systems.

\section{Summary}

We described four separate investigation into the nature of
microlenses and Galactic structure based on the microlensing
events detected by the MACHO collaboration.
Briefly:\\[-0.6cm]
\begin{enumerate}
\item We compared the color-magnitude location of the source stars
for the LMC events with composite CMDs representing different 
lensing scenarios.
If extinction is not too patchy, then it is unlikely that
most of the microlensing events in the LMC have source stars behind
the LMC.
\item We performed a detailed analysis of event LMC-5 from 1993,
which in 1999 was
resolved into 2 stars in the HST images.
We identified one of them as a source and another as a lens.
The lens is a low-mass member of the Galactic disk. Its mass 
$m_{\rm lens}$ is somewhat uncertain, but likely in the range
between $0.04$ and $0.13 \, M_{\odot}$.
\item We described the problem of the microlensing optical depth toward
the Galactic bulge. From clump giant events, we obtained 
$\tau_{\rm bulge} = (2.0 \pm 0.4) \cdot 10^{-6}$
when we averaged over 77 MACHO fields and  
$\tau_{\rm bulge} = (1.4 \pm 0.3) \cdot 10^{-6}$ when
a single highly anomalous field was excluded.
Therefore, we showed that the results from clump giants
are no longer in disagreement with infrared-based 
models of the Galactic bar. The situation is, however, still
complicated due to the fact the Difference Image Analyses result in
substantially higher optical depths. This difference is not
understood.
\item We found several events that show clear signs of parallax
effect in their light curves, and we selected the six strongest
candidates. The average mass of the lenses of these events
is 2.7 solar masses. The two longest events may be
due to black holes with $M_{\rm bh}$ of order of 6 solar masses.
It is very likely that the long-duration events toward the Galactic
bulge are caused by massive remnants.
\end{enumerate}

\acknowledgments

This work was performed under the auspices of the US Department
of Energy, National Nuclear Security Administration, by the University
of California, Lawrence Livermore National Laboratory, under
contract W-7405-ENG-48.
PP thanks the conference organizers for invitation to this exciting
meeting.

\label{page:last}

\end{document}